\documentclass[aps,prb,reprint,showpacs,superscriptaddress,
footinbib,citeautoscript]{revtex4-1}
\usepackage{graphicx}
\usepackage[utf8]{inputenc}
\usepackage{csquotes}
\usepackage[american]{babel}
\usepackage[T1]{fontenc}
\usepackage{enumerate}
\usepackage{mdwlist}
\usepackage[activate=normal]{pdfcprot}
\usepackage{bbding}
\usepackage{color}
\usepackage{amssymb}
\usepackage{amsmath}
\usepackage{amsfonts}
\usepackage{mathrsfs}
\usepackage{bm}
\usepackage{dcolumn}
\usepackage{color}
\usepackage[colorlinks=true,citecolor=blue]{hyperref}

\providecommand{\U}[1]{\protect\rule{.1in}{.1in}}
\hypersetup{colorlinks=true,citecolor=blue,linkcolor=red
,urlcolor=blue}

\begin{document}

\title{Magnetic spheres in microwave cavities}
\author{Babak Zare Rameshti}
\affiliation{Department of Physics, Institute for Advanced Studies in Basic Sciences
(IASBS), Zanjan 45137-66731, Iran}
\affiliation{Kavli Institute of NanoScience, Delft University of Technology, Lorentzweg
1, 2628 CJ Delft, The Netherlands}
\author{Yunshan Cao}
\affiliation{Kavli Institute of NanoScience, Delft University of Technology, Lorentzweg
1, 2628 CJ Delft, The Netherlands}
\author{Gerrit E. W. Bauer}
\affiliation{Kavli Institute of NanoScience, Delft University of Technology, Lorentzweg
1, 2628 CJ Delft, The Netherlands}
\affiliation{Institute for Materials Research and WPI-AIMR, Tohoku University, Sendai
980-8577, Japan}

\begin{abstract}
We apply Mie scattering theory to study the interaction of magnetic spheres
with microwaves in cavities beyond the magnetostatic and rotating wave
approximations. We demonstrate that both strong and ultra-strong coupling
can be realized for a stand alone magnetic spheres made from\ yttrium iron
garnet (YIG), acting as an efficient microwave antenna. The eigenmodes of
YIG spheres with radii of the order mm display distinct higher angular
momentum character that has been observed in experiments.
\end{abstract}

\maketitle

\section{Introduction}

\label{sec:intro}

Light-matter interaction in the strong coupling regime is an important
subject in coherent quantum information transfer.\cite{Wallraff2004,
Kubo2010, Putz2014} Spin ensembles such as nitrogen-vacancy centers may
couple strongly to electromagnetic fields and have the advantage of both
long coherence times \cite{Bar-Gill2013} and fast manipulation \cite%
{Childress2006}. The \textquotedblleft magnon\textquotedblright\ refers to
the collective excitation of spin systems. In paramagnetic spin ensembles in
an applied magnetic field, the spins precess coherently in the presence of
microwave radiation, creating hybridized states referred to as
magnon-polaritons.\cite{Mills1974, Lehmeyer1985, Cao} In the strong coupling
regime coherent energy exchange exceeds the dissipative loss of both
subsystems. The coherent coupled systems is usually described by the
Tavis-Cummings (TC) model \cite{Tavis1968, Fink2009}, which defines a
coupling constant $g$ between the spin-ensemble and the electromagnetic
radiation that scales with the square root of the number of spins. In
ferro/ferrimagnets the net spin density is exceptionally large and
spontaneously ordered, which makes those materials very attractive for
strong-coupling studies. The exchange coupling of spins in magnetic
materials also strongly modifies the excitation spectrum into a spectrum or
spin wave band structure. An ubiquitous experimental technique to study
ferromagnetism is ferromagnetic resonance (FMR), i.e. the absorption,
transmission or reflection spectra of microwaves. In the weak coupling
regime FMR gives direct access to the elementary excitation spectrum of
ferromagnets,\cite{Hillebrands} including the standing spin waves in
confined systems referred to spin wave resonance (SWR).\cite{SWR} The strong
coupling regime is studied less frequently, however, because the dissipative
losses of the magnetization dynamics are usually quite large.

An exceptional magnetic material is the man-made yttrium iron garnets (YIG),
a ferrimagnetic insulator. Commercially produced high-quality spherical YIG samples serve in
magnetically tunable filters and resonators at microwave frequencies. By
suitable doping becomes a versatile class of materials with low dissipation
and unique microwave properties \cite{Serga2010}. YIG has spin density of $%
2\times 10^{22}\mathrm{cm}^{-3}$, \cite{Gilleo1958} and the Gilbert damping
(reciprocal quality) factor of the magnetization dynamics ranges from $%
10^{-5}$ to $10^{-3}$ \cite{Kajiwara2010, Heinrich2011, Kurebayashi2011},
which facilitates strong coupling for smaller samples. Indeed, strongly
coupled microwave photons with magnons have been experimentally reported for
either YIG films with broadband coplanar waveguides (CPWs)\cite{Huebl2013,
Stenning2013, Bhoi2014}, or YIG spheres in 3D microwave cavities \cite%
{Tabuchi2014, Zhang2014, Goryachev2014}. A series of anticrossings were
observed in thicker YIG films and split rings \cite{Stenning2013, Bhoi2014}.
The coupling of magnons in YIG spheres with a superconducting qubit via a
mircowave cavity mode in the quantum limit has been reported \cite%
{Tabuchi2014}. An ultrahigh cooperativity $C=g^{2}/\kappa \gamma >10^{5},$
where $\kappa $ and $\gamma $ are the loss rates of the cavity and spin
system, and multimode strong coupling were found at room \cite{Zhang2014} as
well as the low \cite{Goryachev2014} temperatures.

From a theoretical point of view, the standard TC model is too simple to
describe the full range of coupling between magnets and microwaves. Also the
rotating-wave approximation (RWA) (usually but not necessarily assumed in
the TC model) is speaking applicable when the coupling ratio $g/\omega
_{c}\ll 1$, where $\omega _{c}$ is the microwave cavity mode frequency. We
may define different coupling regimes \cite{Ballester2012, Moroz2014}, viz.
(i) strong coupling (SC) when $0.01<g/\omega _{c}\lesssim 0.1$, (ii)
ultrastrong coupling (USC) \cite{Niemczyk2010} when $g/\omega _{c}\gtrsim 0.1
$, (iii) or even deep strong coupling (DSC) $g/\omega _{c}\approx 1$ \cite%
{Casanova2010}. Cao \textit{et al.} \cite{Cao} adapted the TC model to
ferromagnets by formulating a first-principles scattering theory of the
coupled cavity-ferromagnet system based on the Maxwell and the
Landau-Lifshitz-Gilbert equation including the exchange interaction. A
effectively one-dimensional system of a thin film with in-plane
magnetization in a planar cavity was solved exactly in the linear regime,
exposing, for example, strong-coupling to standing spin waves. Maksymov 
\textit{et al}. \cite{Maksymov} carried out a {numerical study of the
strong coupling regime in all-dielectric magnetic multilayers that
resonantly enhance the microwave magnetic field.} A quantum theory of strong
coupling for nanoscale magnetic spheres in microwave resonators has been
developed in the macrospin approximation \cite{Soykal2010}, but this regime
has not yet been reached in experiments.

Here we apply our classical method \cite{Cao} to spherically symmetric
systems, i.e., a magnetic sphere in the center of a spherical cavity. This is
basically again a one-dimensional problem that can be treated
semi-analytically and has other advantages as well, such as a homogeneous
dipolar field and simple boundary conditions. The eigenmodes of magnetic
spheres have been studied in the \textquotedblleft
magnetostatic\textquotedblright\ approximation \cite{Walker, Fletcher1959}, {%
in which the spins interact by the magnetic dipolar field, disregarding
exchange as well as propagation effects, which may be done when $\lambda \gg
a$, where $a$ is the radius of the sphere and $\lambda $ the wavelength of
the incident radiation}. Arias \textit{et al.} \cite{Arias2005} treated the
interaction of magnetic spheres with microwaves in the weak-coupling regime.
In contrast, we address here the properties of the fully hybridized
magnon-polaritons beyond the magnetostatic approximation (but disregard the
exchange interaction), including the propagation effects (reflection and
transmission) of microwaves, thereby extending the validity to $\lambda <a$.
We are admittedly still one step from the \textquotedblleft
exact\textquotedblright\ solution by disregarding the exchange (as treated
and discussed by Cao \textit{et al.} \cite{Cao}). Our calculated microwave
spectra are complex but help in understanding some of the above-mentioned
experiments.

This manuscript is organized as follows. In Sec.~\ref{sec:model}, we
introduce the details of our model and derive the scattered intensity and
efficiency factors for a strongly coupled system of a magnetic sphere and
microwaves. In Sec.~\ref{sec:results}, we present and discuss our numerical
results that demonstrate the effects both due to the dielectric as well as
magnetic effects on the scattering properties and compare our results with
experiments. In Sec.~\ref{sec:concl}, we conclude and summarize our findings.

\section{Model and formalism}

\label{sec:model}
\begin{figure}[tbp]
\includegraphics[width=8cm]{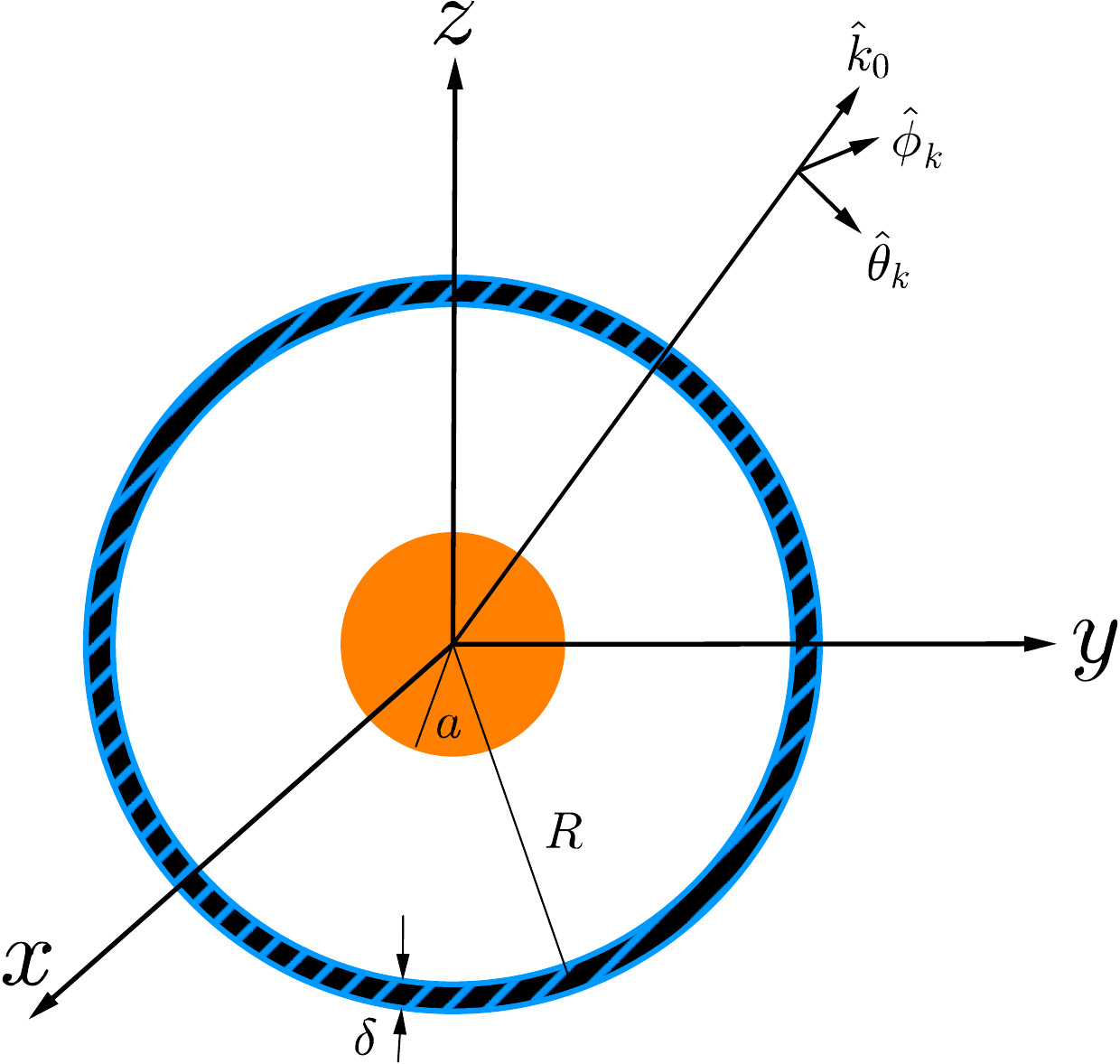}
\caption{(Color online) Plane wave with wave vector $\mathbf{k}_{0}$
coming in at an arbitrary angle hits a large spherical cavity modeled by a
dielectric spherical shell of radius $R$, thickness $\protect\delta $, and
permittivity $\protect\epsilon _{c}$. The spherical cavity is loaded with a
magnetic sphere of radius $a$ centered at the origin of the coordinate
system.}
\label{fig1}
\end{figure}

We model the coupling of the collective excitations of a
magnetic sphere to microwaves in a spherical cavity by the coupled
Landau-Lifshitz-Gilbert and Maxwell equations. We employ Mie-type scattering
theory, i.e., a rapidly converging expansion into spherical harmonics \cite%
{Mie, Kerker, Stratton}. We model the incoming radiation as plane
electromagnetic waves with arbitrary polarization and wave vector that are
scattered by a cavity loaded by a magnetic sphere with gyromagnetic
permeability tensor $\overleftrightarrow{\mu }.$\cite{Gerson1974} In order
to understand the experiments it is not necessary to precisely model the
details of the resonant cavity. Instead, we propose a generic model cavity
that is flexible enough to mimic any realistic situation by adjusting the
parameters. We consider a thin spherical shell of a material with high
dielectric constant $\epsilon _{c}/\epsilon _{0}\gg 1,$ radius $R$, and
thickness $\delta $ that confines standing microwave modes with adjustable
interaction with the microwave source (see Fig. \ref{fig1}). The spherical
symmetry simplifies the mathematical treatment, while the parameters $R$ and 
$\delta $ allow us to freely tune the frequencies and broadenings of the
cavity modes.

The dynamics of the magnetization vector $\mathbf{M}$ is described by the
LLG equation, 
\begin{equation}
\partial _{t}\mathbf{M}=-\gamma \mathbf{M}\times \mathbf{H}_{\mathrm{eff}}+%
\frac{\alpha }{M_{s}}\mathbf{M}\times \partial _{t}\mathbf{M}  \label{eq1}
\end{equation}%
with $\alpha $ and $\gamma $ being the Gilbert damping constant and
gyromagnetic ratio, respectively. The effective magnetic field $\mathbf{H}_{%
\mathrm{eff}}=\mathbf{H}_{\mathrm{ext}}+\mathbf{H}_{\mathrm{x}}$ comprises
the external and (collinear) easy axis anisotropy fields $\mathbf{H}_{%
\mathrm{ext}}$ as well as the exchange field $\mathbf{H}_{\mathrm{x}%
}=J\nabla ^{2}\mathbf{M}$, with $J$ being the exchange stiffness. Assuming
that perturbing microwave magnetic field and magnetization precession angles
are small: 
\begin{align}
\mathbf{M}(\mathbf{r},t)& =\mathbf{M}_{s}+\mathbf{m}(\mathbf{r},t) \\
\mathbf{H}(\mathbf{r},t)& =\mathbf{H}_{\mathrm{ext}}+\mathbf{h}(\mathbf{r},t)
\label{eq2}
\end{align}%
where $\mathbf{M}_{s}$ is the saturated magnetization vector and $\mathbf{m}$
the small-amplitude magnetization driven by the rf magnetic field $\mathbf{h,%
}$ we linearize the LLG equation to 
\begin{equation}
\partial _{t}\mathbf{m}=-\gamma \mathbf{M}_{s}\times \left( \mathbf{H}_{%
\mathrm{eff}}^{(1)}-\frac{\alpha }{\gamma M_{s}}\partial_{t} \mathbf{m}%
\right) -\gamma \mathbf{m}\times \mathbf{H}_{\mathrm{eff}}^{(0)}  \label{eq3}
\end{equation}%
where $\mathbf{H}_{\mathrm{eff}}^{(0)}=\mathbf{H}_{\mathrm{ext}}$ and $%
\mathbf{H}_{\mathrm{eff}}^{(1)}=\mathbf{H}_{\mathrm{x}}+\mathbf{h}$. The
response of ferromagnetic spheres is affected by exchange when their radii
approach the exchange length. Since the latter is typically a few $\mathrm{nm%
}$, we hereafter disregard the exchange interaction and concentrate on the
dipolar spin waves. In the frequency domain and taking the $\mathbf{z}$
direction as the equilibrium direction for the magnetization,
\begin{equation}
i\omega \mathbf{m}=\mathbf{z}\times \left( \omega _{\mathrm{M}}\mathbf{h}%
-\omega _{\mathrm{H}}\mathbf{m}+i\omega \alpha \mathbf{m}\right) ,
\label{eq4}
\end{equation}%
with $\omega _{\mathrm{M}}=\gamma M_{s}$ and $\omega _{\mathrm{H}}=\gamma
H_{0}$. We may recast Eq. (\ref{eq5}) into the form $\mathbf{m}=%
\overleftrightarrow{\chi }\cdot \mathbf{h.}$The magnetic susceptibility
tensor $\overleftrightarrow{\chi }$ is related to the magnetic permeability
tensor by $\overleftrightarrow{\mu }=\mu _{0}(\overleftrightarrow{1}+%
\overleftrightarrow{\chi })$. We find 
\begin{equation}
\overleftrightarrow{\mu }=\mu _{0}%
\begin{pmatrix}
1+\chi & -i\kappa & 0 \\ 
i\kappa & 1+\chi & 0 \\ 
0 & 0 & 1%
\end{pmatrix}
\label{eq5}
\end{equation}%
where $\chi $ and $\kappa $ are given by, 
\begin{align}
\chi & =\frac{(\omega _{\mathrm{H}}-i\alpha \omega )\omega _{\mathrm{M}}}{%
(\omega _{\mathrm{H}}-i\alpha \omega )^{2}-\omega ^{2}}  \label{eq6} \\
\kappa & =\frac{\omega \omega _{\mathrm{M}}}{(\omega _{\mathrm{H}}-i\alpha
\omega )^{2}-\omega ^{2}}.
\end{align}%
The permeability tensor appears in the Maxwell equations for the propagation
of the electromagnetic wave in a magnetic medium.

Inside a spatially homogeneous medium a monochromatic wave with frequency $%
\omega $, 
\begin{align}
\mathbf{\nabla }\times \mathbf{E}& =i\omega \mathbf{b},\quad \mathbf{\nabla }%
\times \mathbf{h}=-i\omega \mathbf{D} \\
\mathbf{\nabla }\cdot \mathbf{D}& =0,\qquad \quad \mathbf{\nabla }\cdot 
\mathbf{b}=0.  \label{eq7}
\end{align}%
The constitutive relation between the magnetic induction $\mathbf{b}$,
electric displacement $\mathbf{D}$, magnetic field $\mathbf{h}$, and the
electric field $\mathbf{E}$ inside this medium are 
\begin{equation}
\mathbf{b}=\overleftrightarrow{\mu }\cdot \mathbf{h},\quad \mathbf{D}%
=\epsilon _{\mathrm{sp}}\mathbf{E}.  \label{eq8}
\end{equation}%
where $\epsilon _{\mathrm{sp}}$ is the scalar permittivity of the medium. It
follows from Eqs. (\ref{eq7}) and (\ref{eq8}) that the magnetic induction $%
\mathbf{b}$ satisfies the wave equation, 
\begin{equation}
\mathbf{\nabla }\times \mathbf{\nabla }\times (\mu _{0}\overleftrightarrow{%
\mu }^{-1}\cdot \mathbf{b})-k_{\mathrm{sp}}^{2}\mathbf{b}=0  \label{eq9}
\end{equation}%
with $k_{\mathrm{sp}}^{2}=\omega ^{2}\epsilon _{\mathrm{sp}}\mu _{0}$.

The surrounding (nonmagnetic) medium is homogeneous and isotropic with
scalar magnetic permeability $\mu _{0}$, divergenceless magnetic field, and
simplified wave equation $\mathbf{\nabla }^{2}\mathbf{b}+k_{\mathrm{sp}}^{2}%
\mathbf{b}=0$. Due to the spherical symmetry it is advantageous to expand
the magnetic field $\mathbf{h}$ into vector spherical harmonics as \cite%
{Kerker,Stratton,ZLin,LWL}
\begin{equation}
\mathbf{h}=\sum_{nm}\bar{\eta}_{nm}\left[ p_{nm}\mathbf{M}_{nm}^{(1)}(k,%
\mathbf{r})+q_{nm}\mathbf{N}_{nm}^{(1)}(k,\mathbf{r})\right] ,  \label{eq10}
\end{equation}%
where $n$ runs from 1 to $\infty $, and $m=-n,\cdots ,n$ with prefactors $%
\bar{\eta}_{nm}=\eta _{nm}k_{0}/(\omega \mu _{0})$, 
\begin{equation}
\eta _{nm}=i^{n}E_{0}\left[ \frac{2n+1}{n(n+1)}\frac{(n-m)!}{(n+m)!}\right]
^{1/2}.
\end{equation}%
$E_{0}$ is the electric field amplitude of the incident wave . The vector
spherical harmonics read \cite{Kerker,Stratton,ZLin,LWL} 
\begin{align}
\mathbf{M}_{nm}^{(j)}(k,\mathbf{r})& =z_{n}^{(j)}(kr)\mathbf{X}_{nm}(\hat{%
\mathbf{r}}),  \notag \\
k\mathbf{N}_{nm}^{(j)}(k,\mathbf{r})& =\mathbf{\nabla }\times \mathbf{M}%
_{nm}^{(j)}(k,\mathbf{r}).  \label{eq12}
\end{align}%
$z_{n}^{(j)}(kr)$ are spherical Bessel functions, $\mathbf{X}_{nm}(\hat{%
\mathbf{r}})=\mathbf{L}Y_{nm}(\hat{\mathbf{r}})/\sqrt{n(n+1)}$ spherical
harmonics and $\mathbf{L}=-i\mathbf{r}\times \mathbf{\nabla }_{\mathbf{r}}$
the angular momentum operator with $\mathbf{\nabla }_{\mathbf{r}}$ the
gradient operator. The electric field distribution is obtained by $\mathbf{E}%
=(i/\omega c)\mathbf{\nabla }\times \mathbf{h}$. By invoking the vector
spherical wave function expansion for $\mathbf{b}$ and $\overleftrightarrow{%
\mu }^{-1}\cdot \mathbf{b}$ in the wave equation Eq. (\ref{eq9}) leads to
the dispersion relation for $k\left( \omega \right) $.

We match the field distributions inside and outside the cavity to obtain the
scattering solution for incident plane microwaves. The field inside the
spherical shell must be regular, while the scattered component has to
satisfy the scattering wave boundary conditions at infinity. These
conditions are fulfilled by adopting the first kind of spherical Bessel
function $j_{n}(x)$ as radial part for the internal distribution and the
first kind of spherical Hankel function $h_{n}^{(1)}(x)$ for the scattered
component outside the cavity 
\begin{equation}
\mathbf{h}_{s}=\sum_{nm}\bar{\eta}_{nm}\left[ c_{nm}\mathbf{N}%
_{nm}^{(3)}(k_{0},\mathbf{r})+d_{nm}\mathbf{M}_{nm}^{(3)}(k_{0},\mathbf{r})%
\right] .  \label{eq14}
\end{equation}%
The unknown scattering coefficients $c_{nm}$ and $d_{nm}$ are determined by
the boundary conditions at the interface. We consider here the situation in
which the magnetic sphere is illuminated by a plane wave with arbitrary
direction of propagation and polarization as indicated in Fig. \ref{fig1}.
The incident field can be expanded as 
\begin{equation}
\mathbf{h}_{inc}=-\sum_{nm}\bar{\eta}_{nm}\left[ u_{nm}\mathbf{N}%
_{nm}^{(1)}(k_{0},\mathbf{r})+v_{nm}\mathbf{M}_{nm}^{(1)}(k_{0},\mathbf{r})%
\right] . \label{eq15}
\end{equation}%
The expansion coefficients $u_{mn}$ and $v_{mn}$,
\begin{align}
u_{nm}& =\left[ p_{\theta }\tilde{\tau}_{nm}(\cos \theta _{k})-ip_{\phi }%
\tilde{\pi}_{nm}(\cos \theta _{k})\right] e^{-im\phi _{k}},  \label{eq16} \\
v_{nm}& =\left[ p_{\theta }\tilde{\pi}_{nm}(\cos \theta _{k})-ip_{\phi }%
\tilde{\tau}_{nm}(\cos \theta _{k})\right] e^{-im\phi _{k}} , \label{eq17}
\end{align}%
contain all information about the polarization vector and direction of
propagation, where $\hat{\mathbf{p}}=(p_{\theta }\hat{\theta}_{k}+p_{\phi }%
\hat{\phi}_{k})$ is the normalized complex polarization vector, with $%
\left\vert \hat{\mathbf{p}}\right\vert =1$ and $\theta _{k}(\phi _{k})$ is
the polar (azimuthal) angle of $\mathbf{k}_{0}$. Two auxiliary functions are
defined by 
\begin{equation}
\tilde{\pi}_{nm}=t_{nm}\frac{m}{\sin \theta }P_{n}^{m}(\cos \theta ),~\tilde{%
\tau}_{nm}=t_{nm}\frac{d}{d\theta }P_{n}^{m}(\cos \theta ),  \label{eq18}
\end{equation}%
with $t_{nm}=i^{-n}\eta _{nm}/E_{0}$ and $P_{n}^{m}(x)$ the first kind
associated Legendre function.

In order to solve the full scattering problem including the cavity we match
the fields outside the cavity caused by the incoming plane microwave and the
spacer region separating the magnetic particle and cavity. In the latter,
spherical Bessel functions of both the first and second kind have to be
included into the expansion. At the surface of the magnetic sphere $\left(
r=a\right) $ we adopt the standard boundary conditions 
\begin{align}
\mathbf{h}_{i}\times \mathbf{e_{r}}& =\mathbf{h}_{mid}\times \mathbf{e_{r}}
\\
\mathbf{\mathbf{E}}_{i}\times \mathbf{e_{r}}& =\mathbf{\mathbf{E}}%
_{mid}\times \mathbf{e_{r}}  \label{eq19}
\end{align}%
while at the surface of the cavity, assuming that its thickness is much
smaller than the wavelength,\cite{MGA-57-1,MGA-57-2} 
\begin{align}
\left[ \mathbf{h}_{mid}-\mathbf{h}_{out}\right] \times \mathbf{e_{r}}& =-\xi %
\left[ \mathbf{e_{r}}\times \mathbf{\mathbf{E}}_{out}\right] \times \mathbf{%
e_{r},} \\
\mathbf{\mathbf{E}}_{mid}\times \mathbf{e_{r}}& =\mathbf{\mathbf{E}}%
_{out}\times \mathbf{e_{r}.}  \label{eq20}
\end{align}%
The indexes \textit{mid} and \textit{out} indicate the regions within and
outside of the cavity, respectively. The unit vector $\mathbf{e_{r}}$ is the
outward normal to the surfaces and $\xi =i\omega (\epsilon _{c}-\epsilon
_{0})\delta $ with permittivity of the cavity shell $\epsilon _{c}$. By
matching the field distributions in the different regions the scattering
coefficients are determined, from which we calculate the observables.
\begin{figure}[tbp]
\includegraphics[width=8.5cm]{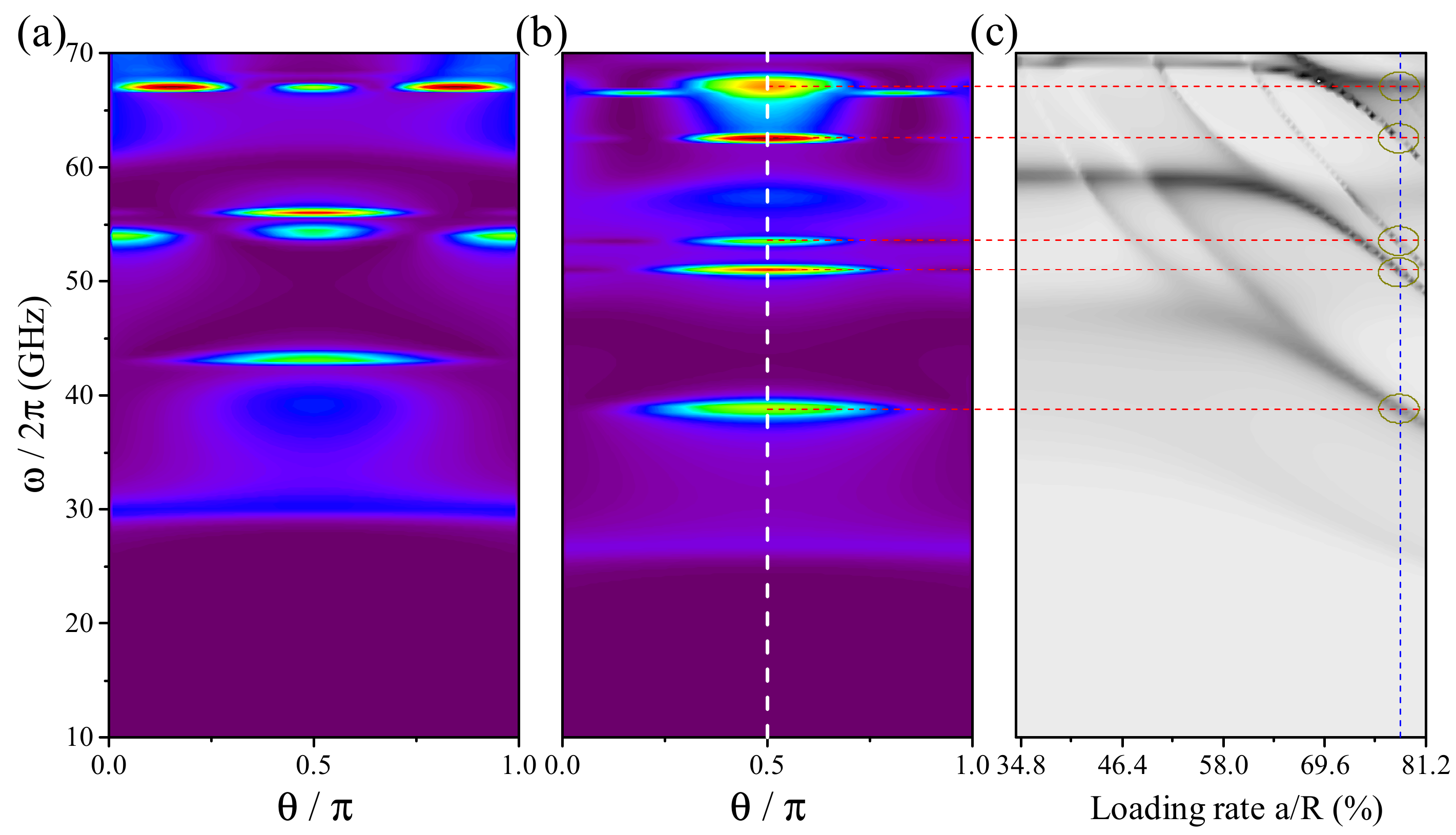}
\caption{(Color online) Scattering intensity $|S_{1}|^{2}$ as function
of scattering angle $\protect\theta $ and frequency $\protect\omega /2%
\protect\pi $ is shown for (a) a dielectric sphere of radius $a=1.25\,%
\mathrm{mm}$ and relative permittivity $\protect\epsilon /\protect\epsilon %
_{0}=15$ and for (b) the same sphere in a cavity of radius $R=1.6\,\,\mathrm{%
mm}$. In (c) the scattering intensity is plotted for the same cavity as
function of frequency and loading rate $a/R$. The dashed lines are guides
for the eye.}
\label{fig2}
\end{figure}
At distances sufficiently far from the cavity, i.e., in the far field zone,
the intensity of the two polarization components $I_{1}$ and $I_{2}$ are 
\begin{align}
I_{1}& \sim \frac{E_{0}^{2}}{k_{0}^{2}r^{2}}|S_{1}(\theta ,\phi )|^{2},
\label{eq21} \\
I_{2}& \sim \frac{E_{0}^{2}}{k_{0}^{2}r^{2}}|S_{2}(\theta ,\phi )|^{2}.
\end{align}%
where $\theta ~(\phi )$ is the polar (azimuthal) angle of the observer at
distance $r$. The scattering amplitude functions are%
\begin{align}
S_{1}(\theta ,\phi )& =\sum_{nm}\left[ d_{nm}\tilde{\tau}_{nm}(\cos \theta
)+c_{nm}\tilde{\pi}_{nm}(\cos \theta )\right] e^{im\phi },  \label{eq27} \\
S_{2}(\theta ,\phi )& =\sum_{nm}\left[ d_{nm}\tilde{\pi}_{nm}(\cos \theta
)+c_{nm}\tilde{\tau}_{nm}(\cos \theta )\right] e^{im\phi },  \label{eq28}
\end{align}%
where the coefficients $c_{nm}$ and $d_{nm}$ characterize the scattered
component of the fields outside the cavity. We may now compute the
scattering and extinction cross sections as well as their (dimensionless)
efficiencies $Q_{sca}$ and $Q_{ext}$, which are the cross sections
normalized by $\pi R^{2}$, the geometrical cross section of the cavity:\ 

\begin{align}
Q_{sca}& =\dfrac{4}{k_{0}^{2}R^{2}}\sum_{nm}\left(
|c_{nm}|^{2}+|d_{nm}|^{2}\right) ,  \label{eq29} \\
Q_{ext}& =\dfrac{4}{k_{0}^{2}R^{2}}\sum_{nm}\mathrm{Re}\left( u_{nm}^{\ast
}d_{nm}+v_{nm}^{\ast }c_{nm}\right) .  \label{eq30}
\end{align}%
The extinction cross section represents the ratio of (angle-integrated)
emitted to incident intensity, i.e., with and without the scattering
cavity/particle between source and detector. This factor measures the energy
loss of the incident beam by absorption and scattering. The series expansion
in Eqs. (\ref{eq27})-(\ref{eq30}) is uniformly convergent and can be truncated
at some point in numerical calculations depending on the desired accuracy.
In the next section we present our results with emphasis on the dielectric
and magnetic contributions to the microwave scattering. 
\begin{figure}[tbp]
\includegraphics[width=8.6cm]{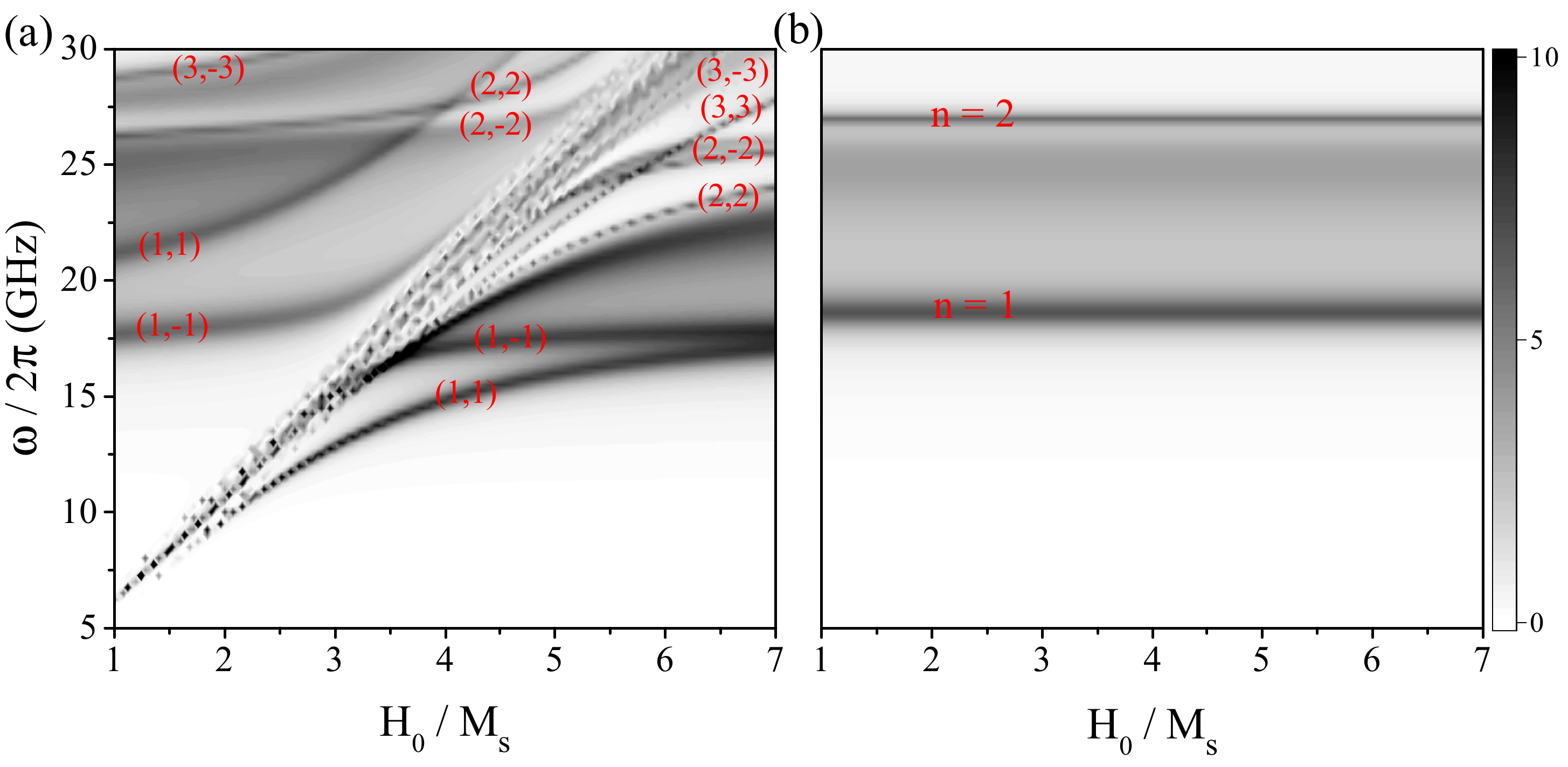}
\caption{(Color online) Panel (a) shows the scattering efficiency factor $%
Q_{sca}$ as function of normalized magnetic field $H_{0}/M_{s}$ and
frequency $\protect\omega /2\protect\pi $ for a YIG sphere of radius $a=2\,%
\mathrm{mm}$ and relative permittivity $\protect\epsilon /\protect\epsilon %
_{0}=15$. Panel (b) shows results for a non-magnetic dielectric sphere. The
character of the microwave modes sufficiently far from the anti-crossing
with the spin waves is labeled by the spherical harmonic indices $(n,m)$.}
\label{fig3}
\end{figure}

\section{Results}

\label{sec:results}Here we present numerical results on the coupling of
microwaves with a ferro- or ferrimagnet in a cavity based on our treatment
of Mie scattering of the electromagnetic waves as exposed in the preceding
section. It applies to a dielectric/magnetic sphere centered in a (larger)
spherical cavity, but both may be of arbitrary diameter otherwise. We are
mainly interested in the coherent coupling between the magnons and microwave
photons in the strong or even ultrastrong coupling regimes that can be
achieved by generating spectrally sharp cavity modes, by increasing the
filling factor of the cavity, or simply by increasing the size of the
sphere. {The RWA, however, tends to break down as the coupling increases.
This has led to different coupling regimes beyond the weak coupling, TC
region, i.e., \textit{strong} (SC) and \textit{ultrastrong} (USC) coupling
regimes. In the SC region coupling strength has to be comparable or larger
than all decoherence rates, while in the USC it has to be comparable or
larger than appreciable fractions of the mode frequency, $g/\omega_{c}
\gtrsim 0.1$.}

We adopt the forward scattered intensities $I_{1}\sim |S_{1}(\theta =\pi
/2,\phi =\pi )|^{2}$ and scattering efficiency factors as convenient and
observable measures of the microwave scattering by a spherical target. In
order to compare results with recent experiments, we chose parameters for
YIG with gyromagnetic ratio $\gamma /(2\pi )=28~$ GHz/T, saturation
magnetization \cite{Grishin} $\mu _{0}M_{s}=175~$ mT, Gilbert damping constant%
\cite{Kajiwara2010, Heinrich2011, Kurebayashi2011} $\alpha =3\times 10^{-4}$, and relative
permittivity \cite{Murthy} $\epsilon /\epsilon _{0}=15.$ Without loss of
generality we consider microwaves incident from the positive $x$ direction ($%
\theta _{k}=\pi /2$ and $\phi _{k}=0$) and polarization {$(p_{\theta
},p_{\phi })=(1,0)$}, so{\ its electric/magnetic components are in the $-$}$%
z/y$ directions (static magnetic field and magnetization $\mathbf{H}%
_{0}\Vert \mathbf{z}$\textbf{)}. Forward scattering is monitored by setting $%
\theta =\pi /2$ and $\phi =\pi $ in Eq. (\ref{eq27}). We also explore the
dependence of the observables on the scattering angles. We can remove the
cavity simply by setting $\xi =0$.

In Fig. \ref{fig2} the scattered intensity $|S_{1}(\theta ,\pi )|^{2}$ is
depicted as a function of frequency $\omega /2\pi $ and scattering angle $%
\theta $ focusing first on a non-magnetic sphere with radius $a=1.25\,%
\mathrm{mm}$. The angular dependence of the scattering with and without a
cavity (with $R=1.6\,\mathrm{mm)}$ is plotted in panel (a) and (b),
respectively. The eigenmodes of the dielectric sphere show $s$, $p$ and $d$%
-wave characters in Fig. \ref{fig2}(a). $s$-wave scattering dominates as
long as the wavelength (reduced by $\epsilon _{\mathrm{sp}}$) does not fit
twice into the sphere, i.e., $\lambda \gtrapprox a\sqrt{\epsilon _{\mathrm{sp}%
}/\epsilon _{0}}$. The spherical cavity, on the other hand, limits the
isotropic scattering regime to $\lambda \gtrapprox R\sqrt{\epsilon _{\mathrm{%
sp}}/\epsilon _{0}}$.

In Fig. \ref{fig2}(c) we plot the forward scattered intensities $I_{1}$%
\textit{\ }as function of the load of the cavity by a dielectric sphere. The
eigenfrequencies of the cavity remain constant, while those confined to the
sphere shift to lower frequencies as $\sim a^{-2}$. At high loading rate the
cavity modes are strongly mixed with the modes in the sphere and all of them
bend towards lower frequencies.

\begin{figure}[tbp]
\includegraphics[width=8.7cm]{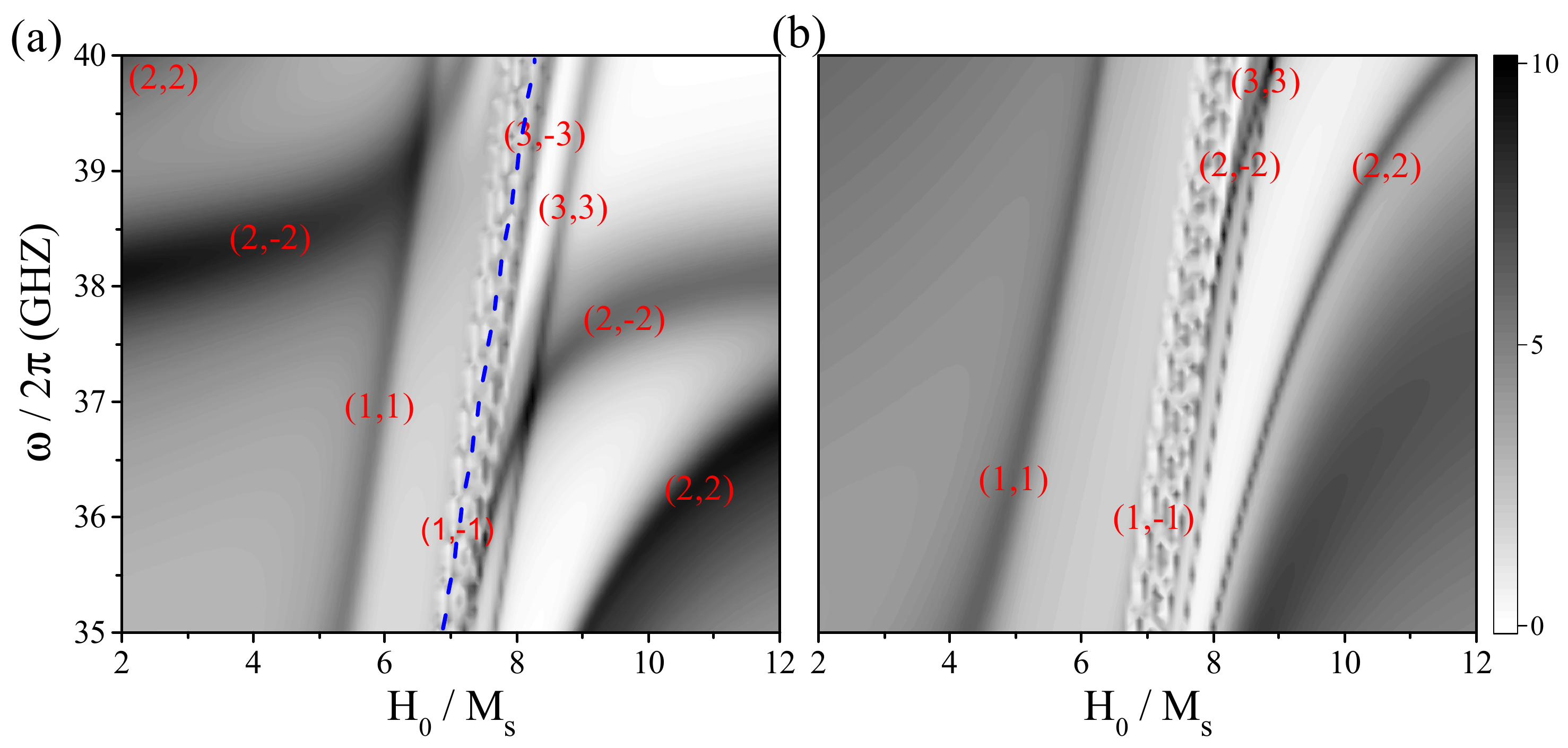}
\caption{(Color online) Scattering efficiency factor $Q_{sca}$ plotted
as function of normalized magnetic field $H_{0}/M_{s}$ and frequency $%
\protect\omega /2\protect\pi $ for a YIG sphere of radius $a=1.25\,\mathrm{mm%
}$ and relative permittivity $\protect\epsilon /\protect\epsilon _{0}=15$
(a) in the center of a spherical cavity of radius $R=1.6\,\mathrm{mm}$ and
(b) without cavity.}
\label{fig4}
\end{figure}

Magnetism of the spheres can affect the microwave scattering properties
strongly, but the issue of hybridization of cavity and sphere resonant
microwave modes is still present. A sufficiently large YIG sphere alone can
therefore provide strong coupling conditions to the magnetization even
without an external resonator. To this end, the linear dimension of the YIG
sphere must be of a size that allows the internal resonances of the sphere
to come into play in the microwave frequency range, i.e. when {$ka\gtrapprox
\pi \sqrt{\epsilon _{0}/\epsilon _{\mathrm{sp}}}$ or $\lambda \lessapprox 2a%
\sqrt{\epsilon _{\mathrm{sp}}/\epsilon _{0}}$}. We therefore have a (narrow)
regime $a\sqrt{\epsilon _{\mathrm{sp}}/\epsilon _{0}}\lessapprox \lambda
\lessapprox 2a\sqrt{\epsilon _{\mathrm{sp}}/\epsilon _{0}}$ or $7.75\,%
\mathrm{mm}\lessapprox \lambda \lessapprox 15.49\,\mathrm{mm}$ (for Fig.\ref%
{fig3}) in which strong coupling and $s$-wave scattering can be realized
simultaneously without a cavity. YIG spheres can typically be fabricated
with high precision for radii in the range\cite{growers} $a=0.9-2.5\,\mathrm{%
mm}$. In Fig \ref{fig3} for a $a=2\,\mathrm{mm}$ YIG sphere we observe a
strong anticrossing between the linear spin wave modes and the
sphere-confined standing microwaves. The YIG sphere is therefore an
efficient microwave antenna that achieves strong and ultra-strong coupling
without a cavity. It should be noted that previous works \cite%
{Filonov, Kuznetsov, Boudarham, Bi, Nikitin}, which have revealed the
possibility to use all-dielectric as well as all-magneto-dielectric
resonators without external resonator, were not considered strong coupling.
\begin{figure}[tbp]
\includegraphics[width=8.7cm]{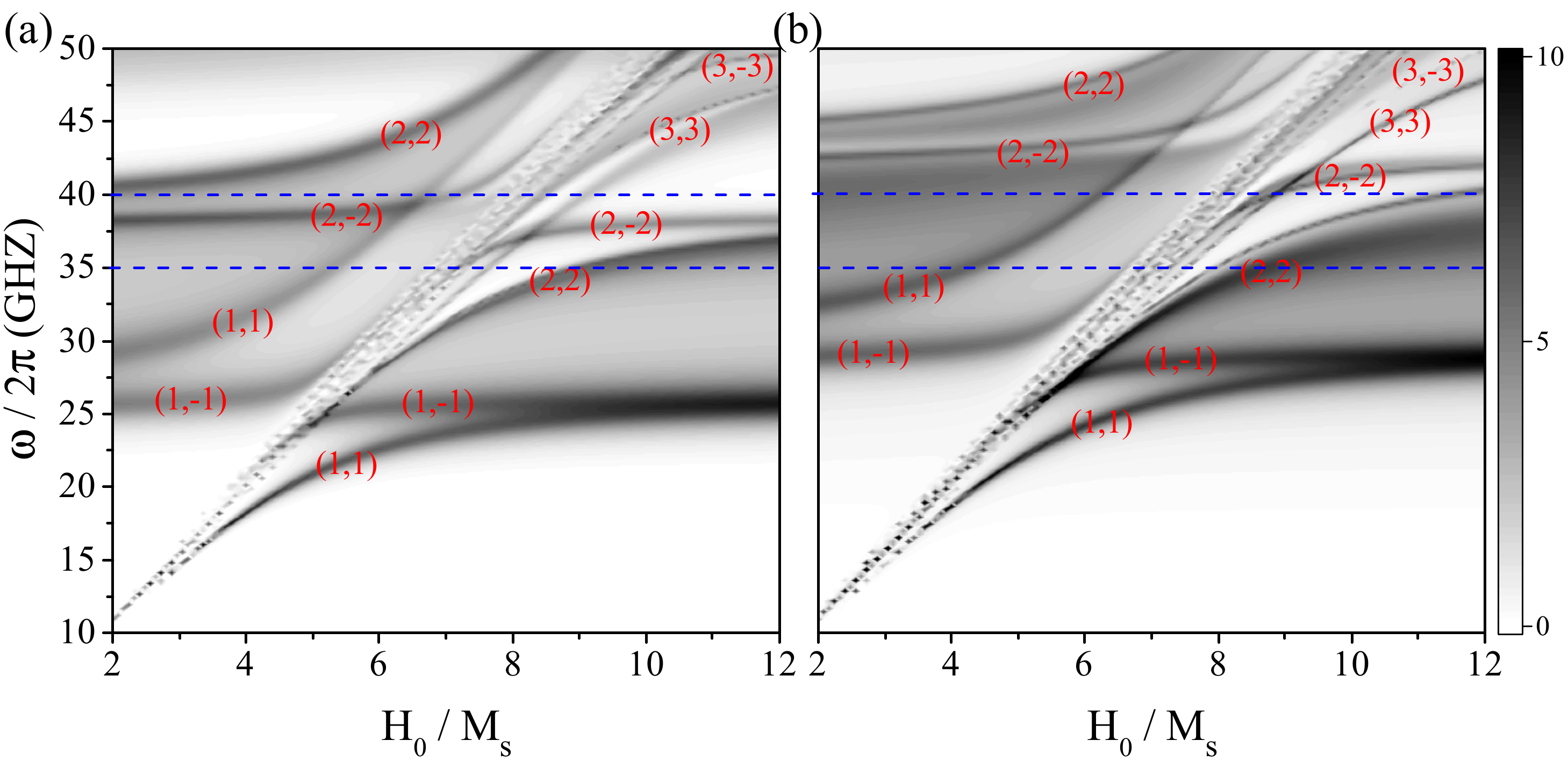}
\caption{(Color online) Scattering efficiency factor $Q_{sca}$ as function
of normalized magnetic field $H_{0}/M_{s}$ and frequency $\protect\omega /2%
\protect\pi $ for a YIG sphere of radius $a=1.25\,\mathrm{mm}$ and relative
permittivity $\protect\epsilon /\protect\epsilon _{0}=15$ (a) in the center
of a spherical cavity of radius $R=1.6\,\mathrm{mm}$ and (b) in the absence
of the cavity. Dashed lines indicate the frequency range in Figure \protect
\ref{fig4}.}
\label{fig5}
\end{figure}

Our results help to interpret recent experimental results on YIG spheres in
microwave cavities with reported coupling strength that are comparable with
the magnon frequency \cite{Zhang2014}, i.e., in the ultrastrong-coupling regime.
In Fig. \ref{fig4} the scattering efficiency factor is shown as a function of $%
H_{0}/M_{s}$ and $\omega /2\pi $. Panel (a) addresses a YIG sphere of radii $%
a=1.25\,\mathrm{mm}$ in a spherical microwave cavity of radii $R=1.6\,%
\mathrm{mm}$, chosen to be close to the leading dimensions of the cavity in
the experiments. Panel (b) holds for the same YIG sphere but without cavity.
The obvious anticrossing in Fig. \ref{fig4}(a) is a signature of the
emergence of the hybrid excitation that we refer to as \textit{%
magnon-polariton}. The anti-crossing modes are labeled by the mode numbers $%
(n,m)$. For given $n$ there are two $m=\pm n$ anti-crossing modes with
coupling strengths $g_{n,n}>g_{n,-n}$, where $g_{n,m}$ is the effective
coupling strength of the magnon mode $(n,m)$ to the cavity. Fig. \ref{fig4}(a) indicates that the ultrastrong-coupling strength is indeed approached
since a splitting of $g/2\pi =2.5~$ GHz is achieved at a resonance frequency
of $\omega /2\pi \simeq 37.5\,$ GHz. Beside the main anticrossing with the 
$(2,2)$ and $(2,-2)$ cavity modes, we observe tails from other
anticrossings with the $(3,3)$ and $(3,-3)$ modes at higher frequencies, as
well as the $(1,1)$ and $(1,-1)$ modes at lower frequencies, which are
standing electromagnetic resonance modes confined by the YIG sphere. We
may interpret these as nearly pure spin wave modes that acquire some
oscillator strengths by mixing from far away resonances due to the
ultrastrong coupling with standing microwaves. This can be verified by
checking the scattering efficiency factor in the absence of the cavity as in
Fig. \ref{fig4}(b), which emphasizes the antenna action of the YIG sphere. 

Zhang \textit{et al.} \cite{Zhang2014} indeed report additional, weakly-coupled
\textquotedblleft higher modes,\textquotedblright but without explaining
their nature. They report ultrastrong-coupling between magnons and the
cavity photons only in the frequency range of $35-40$ GHz, but data at lower
frequencies are not given. In Fig. \ref{fig5} we extend the plots in Fig. %
\ref{fig4} to a larger frequency interval. We observe that the main
anticrossing in the frequency range of $35-40$ GHz is caused by the $n=2$
modes, while hybridized modes originating from the $n=1$ resonance exist at
the lower frequencies. The unperturbed modes between the anticrossing gaps
are therefore not only due to the higher modes, but lower modes with ${n=1}$
also contribute by the ultrastrong-coupling. Two significant curves in the
left and right side of the higher unperturbed modes originate from the
anticrossing modes $n=1$ (the left one) and $n=2$ (the right one) of the YIG
sphere itself, as is more clear in Fig. \ref{fig5}(b) (the computed lines are
broader because we use a relatively large $\kappa $ for computational
convenience). We thereby find again that the strong-coupling
magnon-polariton may form also without cavity. 

We concentrated on the dipolar spin wave excitations
driven by magnetic fields that are strongly inhomogeneous due to a large
dielectric constant. We disregard here exchange interactions, thereby
limiting the validity of the treatment to YIG spheres much larger than the
so-called exchange length that for YIG is only a few nanometers{. In other
words, we cannot properly describe all spin waves with relatively large wave
number or frequencies relatively much higher relative to the FMR frequency}.
Indeed, in the planar configuration spin wave resonances are observable for
rather thick films \cite{Cao}. Exchange-induced whispering gallery modes on
the surface of the YIG might therefore be observable even in thicker
spheres, but their treatment is tedious and beyond the scope of the present
paper.

\section{Conclusion}

\label{sec:concl}

In this paper we implement Mie scattering theory to study the interaction of
dielectric as well as magnetic spheres with microwaves in cavities by the
coupled LLG and Maxwell equations, disregarding only the exchange
interaction. We are mainly interested in the coherent coupling between the
magnons and microwave cavity modes in the strong- or even ultrastrong-coupling regimes characterized by the mode-dependent coupling strengths $%
g_{n,m}$. We reveal that while in the presence of a spherical cavity both
strong and ultrastrong coupling can be realized by tuning the cavity modes
and by increasing the filling factor of the cavity. Surprisingly, these
regimes can also be achieved by removing the external resonator, due to the
strong confinement of electromagnetic waves in sufficiently large YIG
spheres. In this regime, higher angular momentum eigenmodes of the
dielectric sphere participate and the scattering shows $s$- as well as $p$%
-wave character. {We thereby transcend studies that focus on dipolar spin
waves in a magnetostatic framework \cite{Walker, Fletcher1959} by considering
propagation effects via the full Maxwell equation.} Our study might be
useful in designing optimal conditions to design cavities in which YIG
spheres are coherently coupled to, e.g., superconducting qubits, in
microwave cavities for coherent quantum information transfer \cite%
{Tabuchi2014}.

\acknowledgments B.Z.R. thanks S. M. Reza Taheri and Y. M. Blanter for fruitful
discussions. The research leading to these results has received funding from
the European Union Seventh Framework Programme [FP7-People-2012-ITN] under
Grant agreement No. 316657 (Spinicur). It was supported by JSPS Grants-in-Aid
for Scientific Research (Grants No. 25247056, No. 25220910, and No. 26103006), FOM
(Stichting voor Fundamenteel Onderzoek der Materie),\ the ICC-IMR, EU-FET
InSpin 612759, and DFG Priority Programme 1538 \textquotedblleft {%
Spin-Caloric Transport}\textquotedblright\ (BA 2954/1-2).

\end{document}